\PassOptionsToPackage{table}{xcolor}

\documentclass{vldb}
\usepackage{graphicx}
\usepackage{tikz}
\usetikzlibrary{shapes,arrows,shapes.geometric,fit}
\usepackage{caption}
\usepackage{subcaption}
\usepackage{framed}
\usepackage{fancybox}
\usepackage{colortbl}
\usepackage{color}
\usepackage{hyperref}
\usepackage{pgfpages}
\usepackage{amsmath,mathtools,amssymb}
\usepackage{balance}
\usepackage{epigraph}
\usepackage{lineno}

\newtheorem{myex}{Example}

\begin{document}

\title{\huge{$\boldsymbol{\Upsilon}$}\ttlfnt{-DB: Managing Scientific Hypotheses as Uncertain Data}}

\numberofauthors{3} 
\author{
\alignauthor
Bernardo Gon\c{c}alves\\
       \affaddr{LNCC -- National Laboratory for Scientific Computing}\\
       \affaddr{Petr\'opolis, Brazil}\\
       \email{bgonc@lncc.br}
\alignauthor
Fabio Porto\\
       \affaddr{LNCC -- National Laboratory for Scientific Computing}\\
       \affaddr{Petr\'opolis, Brazil}\\
       \email{fporto@lncc.br}
}

\maketitle
\vspace{-15pt}
\begin{abstract}
In view of the paradigm shift that makes science ever more data-driven, $\!$we consider deterministic scientific hypotheses as uncertain data. $\!$This vision comprises a probabilistic data\-base (p-DB) design methodology for the systematic construc\-tion and management of U-relational hypothesis DBs, viz., $\!\Upsilon$-DBs. $\!$It introduces hypothesis management as a promising new class of applications for p-DBs. $\!$We illustrate the potential of $\Upsilon$-DB as a tool for deep predictive analytics.
\end{abstract}

\vspace{-3pt}
\section{Introduction}\label{sec:intro}
\vspace{-9pt}
\setlength{\epigraphrule}{0pt}
\setlength{\epigraphwidth}{.43\textwidth}
\renewcommand{\textflush}{flushepinormal}
\epigraph{
``\emph{Originally, there was just experimental science, and then there was theoretical science, with Kepler's Laws, Newton's Laws of Motion, Maxwell's equations, and so on. Then, for many problems, the theoretical models grew too complicated to solve analytically, and people had to start simulating.}'' 
}{\vspace{-6pt}--- Jim Gray}
\vspace{-6pt}
\noindent
\hspace{-3.4pt}
\noindent
Large-scale experiments provide scientists with \emph{empirical} data that has to be extracted, transformed and loaded before it is ready for analysis \cite{hey2009}. In this vision we consider deterministic scientific hypotheses seen as \emph{theoretical} data, which also needs to be pre-processed to be analyzed, deserving then a proper database approach.

\begin{figure}[hb]
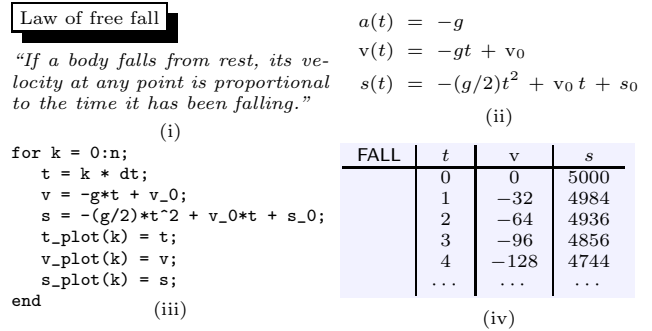

\caption{Multi-fold view of a scientific hypothesis.}
\label{fig:galileo}
\vspace{-8pt}
\end{figure}

$\!\!$\textbf{Hypotheses as data}. As part of the paradigm shift that makes science ever more data-driven, scientific hypotheses are: $\!$(i) formed as principles or ideas, (ii) then mathematically expressed and (iii) implemented as a program that is run to give (iv) their \emph{decisive} form of data (see Fig. $\!$\ref{fig:galileo}).

$\!\!$\textbf{Uncertain data.} $\!$The semantic structure of item (iv) as shown in Fig. \ref{fig:galileo} can be expressed by the functional dependency (FD) $t \to \operatorname{v}\,s$. $\!$This is typical semantics assigned to empirical data in the design of experiment databases. $\!$A space-time dimension (like time $t$ in our example) is used as a key to observables (like velocity $\operatorname{v}$ and position $s$). In \emph{empirical} uncertainty, it is such ``physical'' dimension keys like $t$ that may be violated, say, by alternative sensor readings.

Hypotheses, as tentative explanations of phenomena \cite{losee2001}, are a different kind of uncertain data. In order to manage such \emph{theoretical} uncertainty, we need two special attributes

\noindent\begin{minipage}[t]{0.5\columnwidth}%
\scriptsize{\shadowbox{Law of free fall}}\vspace{1pt}\\
\scriptsize{\emph{``If a body falls from rest, its velocity at any point is proportional to the time it has been falling.''}
\begin{center}
\vspace{-4.5pt}
(i)
\end{center}
}
\vspace{-16.5pt}
\begin{scriptsize}
\begin{verbatim}
for k = 0:n;
   t = k * dt; 
   v = -g*t + v_0; 
   s = -(g/2)*t^2 + v_0*t + s_0; 
   t_plot(k) = t; 
   v_plot(k) = v; 
   s_plot(k) = s;
end
\end{verbatim}
\end{scriptsize}
\begin{center}
\vspace{-13.5pt}
(iii)
\end{center}
\end{minipage}
\hspace{0.2pt}	
\noindent\begin{minipage}[t]{0.5\columnwidth}%
\vspace{-20pt}
\begin{scriptsize}
\begin{eqnarray*}
a(t) \!\!\!&=&\!\!\! -g\\
\operatorname{v}(t) \!\!\!&=&\!\!\! -g t \,+\, \operatorname{v_0}\\
s(t) \!\!\!&=&\!\!\!  -(g/2)t^2 \,+\, \operatorname{v_0}t \,+\, s_0
\end{eqnarray*}
\begin{center}
\vspace{-2.5pt}
(ii)
\end{center}
\vspace{-2pt}
\begingroup\setlength{\fboxsep}{1pt}
\colorbox{blue!5}{%
   \begin{tabular}{c|c|c|c}
  \textsf{FALL} & $t$ & $\operatorname{v}$ & $s$\\
      \hline    
  & $0$ & $0$ & $5000$\\
  & $1$ & $-32$ & $4984$\\
  & $2$ & $-64$ & $4936$\\
  & $3$ & $-96$ & $4856$\\
  & $4$ & $-128$ & $4744$\\
  & $\cdots$ & $\cdots$ & $\cdots$\\
   \end{tabular}
}\endgroup
\begin{center}
\vspace{-3.5pt}
(iv)
\end{center}
\end{scriptsize}
\end{minipage}
\vspace{-8pt}

\vspace{-12pt}
\noindent
to compose, say, the epistemological dimension of keys to observables: $\phi$, identifying the studied phenomena; and $\upsilon$, identifying the hypotheses aimed at explaining them. That is, we shall leverage the semantics of \mbox{item (iv) to $\phi\, \upsilon\, t \to \operatorname{v} s$}. This leap is a core abstraction in the $\Upsilon$-DB vision (see \S \ref{sec:encoding}).

$\!$\textbf{Predictive data.} \mbox{Scientific hypotheses are tested by $\!$way} of their predictions \cite{losee2001}. In the form of mathematical equations, hypotheses symmetrically relate aspects of the studied phenomenon. However, for computing predictions, deterministic hypotheses are used asymmetrically as \emph{functions} \cite{simon1966}. 
$\!$They take a given valuation over input variables (parameters) to produce values of output variables (the predictions). By observing that, in \S \ref{sec:encoding} we introduce a method to extract the FD schema of a hypothesis from its equations.

\textbf{Big data.} Scientific hypotheses qualify to at least four of the five v's associated to the notion of big data: \emph{veracity}, due to their uncertainty; \emph{value}, because of their role in advancing science and technology; \emph{variety}, due to their structural heterogeneity (as noticeable in their FD schemes); and \emph{volume}, because of the large scale of modern scientific problems. 

\textbf{Applications.} $\!$Computational Science research programs such as the Human Brain Project or Cardiovascular$\!$ Mathematics are highly-demanding applications challenged by such theoretical big data. Users need to analyze results of thousands of data-intensive simulation trials.
Also, there is a pressing call for \emph{deep} predictive analytic tools to support users assessing what-if scenarios in business enterprises \cite{haas2011}. All that motivates why \emph{hypothesis management} is a promising class of applications for probabilistic databases (p-DBs).

However, despite the advanced state of the art of probabilistic data management techniques, a lack of systematic methods for the design of p-DBs may prevent wider adoption. In analogy with the field of Graphical Models (GM), considered to inform research in p-DBs \cite[p. $\!$14]{suciu2011}, one of the key success factors for the rapid growth of applications was the availability of systematic methods of construction $\!$\cite{darwiche2010}. The vision of \mbox{$\Upsilon$-DB} addresses that gap by bringing forward one such methodology on top of U-relations and probabilistic world-set algebra (p-$\!$WSA) \cite{koch2009}. 

\vspace{1pt}
\textbf{Predictive analytics.} 
Deep predictive analytics \cite{haas2011} is meant to support users in assessing the consequences of alternative hypotheses. If these can be identified (see \S \ref{sec:encoding}), and their uncertainty is quantified by some probability distribution (see \S \ref{sec:transformation}), then they can be ranked and browsed by the user under selectivity criteria. Furthermore, their probabilities can be conditioned on observed data such that they are possibly re-ranked in the presence of evidence (see \S \ref{sec:analytics}).

\vspace{-6pt}
\section{Related Work}\label{sec:related-work}

\noindent
Haas et al. \cite{haas2011} propose a long-term \emph{models-and-data} research program to address deep predictive analytics. 
Our vision, with roots in \cite{porto2011}, is essentially an abstraction of \emph{hypotheses as data}. It can be understood in comparison as putting models strictly into a data perspective. $\!$Thus it is directly applicable by building upon recent work on p-DBs \cite{suciu2011}. Our vision comprises a \emph{p-DB design meth\-odology} for the systematic construction and management of $\,$U-relational hypothesis DBs, $\!$viz., $\!\Upsilon$-DBs. $\!$It applies classical FD theory \cite{ullman1988} and the U-relatio\-nal representation system with its p-$\!$WSA query algebra \cite{koch2009}. It is not to be confused, say, with initiatives to revisit FD theory in view of uncertain DB design \cite{sarma2007}.

U-relations and p-$\!$WSA were developed in the influential \textsf{MayBMS} project.\footnote{\url{http://maybms.sourceforge.net/}} As implied by its design principles, e.g., compositionality and the ability to introduce uncertainty, \textsf{MayBMS}' query language $\!$\cite{koch2009} fits well to hypothesis management. Noteworthy, the \texttt{repair $\!$key} operation gives rise to alternative worlds as maximal-subset repairs of an argument key. We shall look at it from the point of view of p-DB design, for which no methodology has yet been proposed. 

Again in analogy with GMs, it may be clarifying to distinguish methods for p-DB design in three classes \cite{darwiche2010}: (i) \emph{subjective} construction, (ii) \emph{synthesis} from other kind of formal specification, and (iii) \emph{learning} from data. The first is the less systematic, as the user has to model for the data and correlations by steering all the p-DB construction process (\textsf{MayBMS}' use cases \cite{koch2009}, e.g., are illustrated that way). The second is typified by the \mbox{$\Upsilon$-DB} methodology, as we extract FDs and synthesize U-relations from mathematical equations as a kind of formal specification (see \S\ref{sec:encoding}-\S\ref{sec:transformation}). To our knowledge, this is the first work to propose a synthesis method for p-DB design. The third comprises analytical techniques to extract the data and model correlations from external sources, possibly unstructured, into a p-DB. This is the prevalent one, motivated by information extraction, data integration and data cleaning applications \cite[p.$\!$ 10-3]{suciu2011}. 

Also related to the \mbox{$\Upsilon$-DB} vision is the topic of conditioning a p-DB. It has been firstly addressed by Koch and Olteanu motivated by data cleaning applications \cite{koch2008}. They have introduced the \texttt{assert} operation to implement, as in AI, a kind of knowledge compilation, viz., world elimination in face of constraints (e.g., FDs). For hypothesis management, nonetheless, we need to apply \emph{Bayes' conditioning} by asserting observed data (not constraints). In $\!$\S \ref{sec:analytics}, we present an example that settles the kind of conditioning problem that is relevant to \mbox{$\Upsilon$-DB}. 

In order to provide a concrete feel of our vision, in the next sections we present preliminary results on our methodology for constructing an \mbox{$\Upsilon$-DB} on top of \textsf{MayBMS}.

\begin{figure}[t]\scriptsize
   \centering
\begingroup\setlength{\fboxsep}{5pt}
\colorbox{blue!4}{%
   \begin{tabular}{c|c|p{0.48\linewidth}}
  \textsf{PHENOMENON} & $\phi$ & \textsf{Description}\\
      \hline    
   & $1$ & Effects of gravity on an object falling in the Earth's atmosphere.\\
   \end{tabular}
}\endgroup
\vspace{1pt}
\begingroup\setlength{\fboxsep}{5pt}
\colorbox{blue!4}{%
   \begin{tabular}{c|c|l}
  \textsf{HYPOTHESIS} & $\upsilon$ & \textsf{Name}\\
      \hline    
   & $1$ & Law of free fall\\
   & $2$ & Stokes' law\\
   & $3$ & Velocity-squared law\\
   \end{tabular}
}\endgroup
\vspace{-8.5pt}
\caption{Descriptive (textual) data of Example \ref{ex:fall}.}
\label{fig:research}
\vspace{-7pt}
\end{figure}

\vspace{2pt}
\section{Hypothesis Encoding}\label{sec:encoding}
\noindent
Let us consider Example \ref{ex:fall} to illustrate our methodology.
\vspace{-1pt}
\begin{myex}\label{ex:fall}
$\!$A research is conducted on the effects of gravity on a falling object in the Earth's atmosphere. Scientists are uncertain about the precise object's density and its predominant state as a fluid or a solid. Three hypotheses are then considered as alternative explanations of the fall (see Fig. \ref{fig:research}). $\!$Because of parameter uncertainty, six simulation trials are run for \emph{\textsf{H}}$_1$, and four for \emph{\textsf{H}}$_2$ and \emph{\textsf{H}}$_3$ each. $\Box$
\end{myex}
\vspace{-2pt}
\noindent
The construction of $\Upsilon$-DB requires a simple user description of a research. \mbox{Hypotheses must be associated to the pheno}\-mena they explain and then assigned a prior confidence distribution which may or may not be uniform (see Fig. $\!$\ref{fig:lineage}, $\!$top). $\!$Then the FD schema of each hypothesis has to be extracted from its mathematical equations. Let us examine (\textsf{H$_1$}) the \emph{law of free fall} (Fig. \ref{fig:galileo}) and its set $\Sigma_{1}$ of FDs. 

\vspace{-9pt}
\begin{footnotesize}
 \begin{eqnarray*}\label{sigma1}
\Sigma_{1} = \{\quad 
\phi &\to& g\,\operatorname{v_0}\,s_0,\\ 
g\,\upsilon &\to& a,\\
g\,\operatorname{v_0}\,t\,\upsilon &\to& \operatorname{v},\\
g\,\operatorname{v_0}\,s_0\,t\,\upsilon &\to& s \quad\}.
\end{eqnarray*}
\end{footnotesize}

\vspace{-12pt}
\noindent
In order to derive $\Sigma_{1}$ from the equations in \mbox{Fig. \ref{fig:galileo}-(ii)}, we focus on their hidden data dependencies and get rid of constants and possibly complex mathematical constructs. Equation $\operatorname{v}(t) \!=\! -gt \!+\! \operatorname{v_0}$, e.g., written this way, roughly speaking, allows us to infer that $\operatorname{v}$ is a prediction variable \emph{functionally dependent} on $t$ (the physical dimension), $g$ and $\operatorname{v_0}$ (the parameters). Yet a dependency like $\,g\operatorname{v_0} t \to \operatorname{v}\,$ may hold for infinitely many equations. We need a way to precisely identify \textsf{H$_1$}'s formulation, i.e., an abstraction of its data-level semantics. This is achieved by introducing hypothesis id $\upsilon$ as a special attribute in the FD (see $\Sigma_{1}$). This is a \emph{data representation} of a scientific hypothesis. The other special attribute, the phenomenon id $\phi$, is supposed to be a key to the value of parameters, i.e., determination of parameter values is an empirical, phenomenon-dependent task. FD $\phi \to g\operatorname{v_0} s_0\,$ is to be (expectedly) violated when the user is uncertain about the values of parameters.

\begin{figure*}[t]\scriptsize
\centering
\begin{subfigure}{0.4\linewidth}
\centering
\begingroup\setlength{\fboxsep}{3pt}
\colorbox{blue!5}{
   \begin{tabular}{c|c|c|c}
  \textsf{EXPLANATION} & $\phi$ & $\upsilon$ & \textsf{Conf}\\
      \hline    
   & $1$ & $1$ & $0.6$\\
   & $1$ & $2$ & $0.2$\\
   & $1$ & $3$ & $0.2$\\
   \end{tabular}
}\endgroup\\
\vspace{2pt}
\begingroup\setlength{\fboxsep}{3pt}
\colorbox{blue!5}{%
   \begin{tabular}{c|c|c|c|c|c}
  \textsf{H1\_INPUT} & \textsf{tid} & $\phi$ & $g$ & $\operatorname{v_0}$ & $s_0$\\
      \hline    
   & $1$ & $1$ & $32$ & $0$ & $5000$\\
   & $2$ & $1$ & $32$ & $10$ & $5000$\\
   & $3$ & $1$ & $32$ & $20$ & $5000$\\
   & $4$ & $1$ & $32.2$ & $0$ & $5000$\\
   & $5$ & $1$ & $32.2$ & $10$ & $5000$\\
   & $6$ & $1$ & $32.2$ & $20$ & $5000$\\
   \end{tabular}
}\endgroup\\
\vspace{2pt}
\begingroup\setlength{\fboxsep}{3pt}
\colorbox{blue!5}{%
   \begin{tabular}{c|c|c|c|c}
  \textsf{H1\_OUTPUT[$a$]} & \textsf{tid} & $\phi$ & $\upsilon$ & $a$\\
      \hline    
   & $1$ & $1$ & $1$ & $-32$\\
   & $2$ &  $1$ & $1$ & $-32$\\
   & $3$ &  $1$ & $1$ & $-32$\\
   & $4$ &  $1$ & $1$ & $-32.2$\\
   & $5$ &  $1$ & $1$ & $-32.2$\\
   & $6$ &  $1$ & $1$ & $-32.2$\\
   \end{tabular}
}\endgroup
\caption{Simulation raw data: $\!$trials on \textsf{H}$_1\!$ identified by \textsf{tid}.}\label{fig:lineage}
\end{subfigure}
\textcolor{red}{$\rightarrow$}
\begin{subfigure}{0.5\linewidth}
\centering
\begingroup\setlength{\fboxsep}{3pt}
\colorbox{yellow!15}{%
   \begin{tabular}{c|c|>{\columncolor[gray]{0.97}}c}
  $W$ & $V \mapsto D$ & \textcolor{red}{\textsf{Pr}}\\
      \hline    
   & $x_1 \mapsto 1$ & $.6$\\
   & $x_1 \mapsto 2$ & $.2$\\
   & $x_1 \mapsto 3$ & $.2$\\   
\cline{2-3} 
   & $x_2 \mapsto 1$ & $.5$\\
   & $x_2 \mapsto 2$ & $.5$\\
   \end{tabular}
}\endgroup
\vspace{1pt}\\
\begingroup\setlength{\fboxsep}{3pt}
\colorbox{yellow!12}{%
   \begin{tabular}{c|c||c|c}
  \textsf{Y[Exp]} & $V \mapsto D$ & $\phi$ & $\upsilon$\\
      \hline    
   & $x_1 \mapsto 1$ & $1$ & $1$\\
   & $x_1 \mapsto 2$ & $1$ & $2$\\
   & $x_1 \mapsto 3$ & $1$ & $3$\\
   \end{tabular}
}\endgroup
\vspace{1pt}
\begingroup\setlength{\fboxsep}{3pt}
\colorbox{yellow!12}{%
   \begin{tabular}{c|c||c|c}
  \textsf{Y1[$g$]} & $V \mapsto D$ & $\phi$ & $g$\\
      \hline    
   & $x_2 \mapsto 1$ & $1$ & $32$\\
   & $x_2 \mapsto 2$ & $1$ & $32.2$\\
   \end{tabular}
}\endgroup
\vspace{1pt}
\hspace{.25pt}
\begingroup\setlength{\fboxsep}{3pt}
\colorbox{yellow!12}{%
   \begin{tabular}{c|c|c||c|c|c}
  \textsf{Y1[$a$]} & $V_1 \mapsto D_1$ & $V_2 \mapsto D_2$ & $\phi$ & $\upsilon$ & $a$\\
      \hline    
   & $x_1 \mapsto 1$ & $x_2 \mapsto 1$ & $1$ & $1$ & $-32$\\
   & $x_1 \mapsto 1$ & $x_2 \mapsto 2$ & $1$ & $1$ & $-32.2$\\
   \end{tabular}
}\endgroup
\vspace{1pt}
\begingroup\setlength{\fboxsep}{3pt}
\colorbox{yellow!12}{%
   \begin{tabular}{c|c|c||c|c|c}
  \textsf{Y[$a$]} & $V_1 \mapsto D_1$ & $V_2 \mapsto D_2$ & $\phi$ & $\upsilon$ & $a$\\
      \hline    
   & $x_1 \mapsto 1$ & $x_2 \mapsto 1$ & $1$ & $1$ & $-32$\\
   & $x_1 \mapsto 1$ & $x_2 \mapsto 2$ & $1$ & $1$ & $-32.2$\\
   & $x_1 \mapsto 2$ & $-$ & $1$ & $2$ & $0$\\
   & $x_1 \mapsto 3$ & $-$ & $1$ & $3$ & $0$\\
   \end{tabular}
}\endgroup
\caption{Probabilistic $\Upsilon$-DB storing \textsf{H}$_i$ as $\Upsilon_i$, for $i=1..3$, in \textsf{MayBMS}.}\label{fig:probabilistic}
\end{subfigure}
\vspace{-7pt}
\caption{Synthesis of probabilistic $\Upsilon$-DB from FD schemes and the simulation input/output data ($\Upsilon$-DB's raw data).}\label{fig:db-synthesis}
\vspace{-11pt}
\end{figure*}

The same rationale applies to derive $\Sigma_{2} \!=\! \Sigma_{3}\!$ from the equa\-tions of \textsf{H}$_2$, \textsf{H}$_3\!$ below. $\!$These vary in structure w.r.t. \textsf{H}$_1$ (e.g., parameter $D$, the object's diameter). The key point here is that the method to extract the hypothesis FD schema from its equations is reducible to a lan\-guage for mathematical modeling (based on W3C's \textsf{MathML}).

\vspace{3pt}
\begin{scriptsize}\noindent
\begingroup\setlength{\fboxsep}{2pt}
\colorbox{gray!7}{%
\begin{tabular}{l | l}
\advance\leftskip-3cm
$\!\textsf{H}_2$. \textsf{Stokes' law} & $\!\textsf{H}_3$. \textsf{Velocity-squared law}\\
\hline
$\!a(t) \!= 0$ & $\!a(t) \!= 0$\\
$\!\operatorname{v}(t) \!=\! -\sqrt{gD/\,4.6\!\times\! 10^{-4}}$ & $\!\operatorname{v}(t) \!=\! -gD^2/\,3.29\!\times\! 10^{-6}$\\
$\!s(t) \!=\!  -t\,\sqrt{gD/\,4.6\!\times\! 10^{-4}} \!+\! s_0\!$ & $\!s(t) \!=\!  -(gD^2/\,3.29\!\times\! 10^{-6})\,t \!+\! s_0\!\!\!\!$\\
\end{tabular}
}\endgroup
\end{scriptsize}
\vspace{-3pt}
\begin{footnotesize}
 \begin{eqnarray*}
\Sigma_{2} = \Sigma_{3} = \{\quad 
\phi &\to& g\,D\,s_0,\\ 
\upsilon &\to& a,\\
g\,D\,\upsilon &\to& \operatorname{v},\\
g\,D\,s_0\,t\,\upsilon &\to& s \quad\}.
\end{eqnarray*}
\end{footnotesize}

\vspace{-13pt}
$\!$Once each hypothesis FD schema has been extracted, some reasoning is to be performed to synthesize its certain relations. $\!$The decomposition and pseudo-transiti\-vity inference rules $\!$\cite{ullman1988} $\!$on $\{\phi \to g\operatorname{v_0} s_0,\,g\, \upsilon \to a\} \subset \Sigma_1$, e.g., give $\phi\, \upsilon \to a\,$. Yet there is an extra attribute, \textsf{tid}, added by default to such relations (see Fig. $\!$\ref{fig:lineage}) in order to identify each simulation trial and ``pretend'' completeness.\footnote{Considering $\{\textsf{tid},\,\phi\}$ and $\{\textsf{tid},\,\phi,\,\upsilon\}$ as keys in the relations.} It is under this completeness that the ``raw'' data is loaded from \mbox{input/output} simulation files. Note, however, that it is held at the expense of redundancy and, mostly important, opaqueness for predictive analytics (since \textsf{tid} isolates or hides inconsistency). This is until the next stage of the $\Upsilon$-DB construction method, when the uncertainty is introduced in a controlled manner.

\vspace{-5pt}
\section{Uncertainty Introduction}\label{sec:transformation}
\noindent
The transformation of relations in Fig. \ref{fig:lineage} to probabilistic $\Upsilon$-DB starts with query \textsf{Q}$_1$, creating relation \textsf{Y[Exp]} \mbox{(Fig. \ref{fig:probabilistic})}.

\vspace{3pt}
\begin{scriptsize}
\textsf{Q$_1$. \textbf{create table} Y\_Exp \textbf{as select} phi, upsilon \textbf{from}}\\
\indent\indent \textsf{(\textbf{repair key} phi \textbf{in} EXPLANATION \textbf{weight by} Conf);}
\end{scriptsize}
\vspace{5pt}

\noindent
The $Y$-relations (Fig. \ref{fig:probabilistic}) have in their schema a set of pairs $(V_i, D_i)$ of \emph{condition columns} (cf.$\!$ \cite{koch2009}) to map each discrete random variable $x_i$ to one of its possible values (e.g., $x_1 \!\mapsto\! 1$). The world table $W\!$ stores their marginal probabilities. 

$\!$We create decompositions \textsf{Y1[$\vec{X}$]} for each independent uncertainty unit $\vec{X}\!\subseteq\! \vec{A}$ in \textsf{H1\!\_INPUT(tid, $\!\phi, \vec{A}$)}. $\!$Query \textsf{Q}$_2$, e.g., maps the possible values of $g$ to random variable $x_2$.

\vspace{4pt}
\begin{scriptsize}
\textsf{Q$_2$. \textbf{create table} Y1\_g \textbf{as select} U.phi, U.g \textbf{from}}\\
\indent\indent \textsf{(\textbf{repair key} phi \textbf{in} (\textbf{select} phi, g, \textbf{count(*)} \textbf{as} Fr \textbf{from}}\\ 
\indent\indent \textsf{H1\_INPUT \textbf{group by} phi, g) \textbf{weight by} Fr) as U;}
\end{scriptsize}
\vspace{6pt}

\noindent
The result set of \textsf{Q$_2$} is stored in \textsf{Y1[$g$]} (see Fig. \ref{fig:probabilistic}). Note that we consider relation \textsf{H1\_INPUT} as a joint probability distribution on the values of \textsf{H}$_1$'s parameters and it may not be uniform: we count the frequency \textsf{Fr} of each possible value of an uncertainty factor $\vec{X}\!\subseteq\! \vec{A}$ (as done for $g$ in \textsf{Q$_2$}) and pass it as  argument to the \texttt{weight by} construct.

Then, by considering $\,g\,\upsilon \!\to\! a \in \Sigma_1$, we are able to synthesize prediction relation \textsf{Y1[$a$]} as a query: since $a$ is functionally determined by $\upsilon$ and $g$ only, and these are independent, we propagate their uncertainties onto $a$ by query \textsf{Q$_3$} in the local scope of \textsf{Y1[$a$]} (and similarly for \textsf{Y2[$a$]} and \textsf{Y3[$a$]}).  

\vspace{2pt}
\begin{scriptsize}
\textsf{Q$_3$. \textbf{create table} Y1\_a \textbf{as select} H.phi, H.upsilon, H.a \textbf{from}}\\
\indent\indent \textsf{H1\_OUTPUT\_a \textbf{as} H, Y\_Exp \textbf{as} E, 
Y1\_g \textbf{as} G, (\textbf{select} \textbf{min}(tid)}\\
\indent\indent \textsf{\textbf{as} tid, phi, g \textbf{from} H1\_INPUT \textbf{group by} phi, g) \textbf{as} U}\\
\indent\indent \textsf{\textbf{where} H.tid=U.tid \textbf{and} G.phi=U.phi \textbf{and} G.g=U.g}\\
\indent\indent \textsf{\textbf{and} H.phi=E.phi \textbf{and} H.upsilon=E.upsilon;}
\end{scriptsize}
\vspace{4pt}

\noindent
Query \textsf{Q$_3^\prime$} (not shown) is a \emph{union $\!$all} query selecting $\phi$, $\upsilon$ and $a$ from \mbox{\textsf{Yi[$a$]}} for each $i\!=\!1..3$. $\!\!$The result sets of \textsf{Q$_3$} and \textsf{Q$_3^\prime$}, resp. \textsf{Y1[$a$]} and \textsf{Y[$a$]}, are shown in Fig. \ref{fig:probabilistic}. 

Now, compare relations \textsf{H1\_OUTPUT[$a$]} and \textsf{Y1[$a$]}. 
By accounting for the correlations captured in the FD $\,g\,\upsilon \to a$, we could propagate onto $a$ the uncertainty coming from the hypothesis and the only parameter it is sensible to, thus precisely situating tuples of \textsf{Y1[$a$]} in the space of possible worlds. $\!$The same is done for predictive attributes $\operatorname{v}$ and $s$. In the end, we have $\Upsilon$-DB ready for predictive analytics, i.e., with all competing predictions as possible alternatives which are mutually inconsistent.

The key point here is that all the synthesis process is ame\-nable to algorithm design. Except for the user research description, the $\Upsilon$-DB construction is fully automated based on the FD schemes and the simulation raw data.

\vspace{-6pt}
\section{Predictive Analytics}\label{sec:analytics}

\noindent
Users of Ex. $\!$\ref{ex:fall}, has to be able, say, to query \mbox{phenomenon $\phi\!=\!1$} w.r.t. predicted position $s$ at specific times $t$ by considering all hypotheses $\upsilon$ admitted. That is illustrated by query \textsf{Q$_4$}, which creates integrative table \textsf{Y[$s$]}; and by query \textsf{Q$_5$}, which computes the confidence aggregate \cite{koch2009} for all $s$ tuples where $t\!=\!3$ (Fig. \ref{fig:analytics} shows \textsf{Q$_5$}'s result, apart from column \textsf{Posterior}).

The confidence on each hypothesis for the specific prediction of \textsf{Q}$_5$ is split due to parameter uncertainty such that they sum up back to its \mbox{total confidence. $\!$For \textsf{H}$_2\!$ and \textsf{H}$_3$, $\!$e.g.,} we have $\{g\,D\,s_0\,t\,\upsilon \!\to\! s\} \,\subset\, \Sigma_2 \!=\! \Sigma_3$. Since $g$ and $D$ are the parameter uncertainty factors of $s$ ($s_0$ is certain), with 2 possible values (not shown) each, $\,$then there$\,$ are$\,$ only $2\!\times\! 2\!=\!4\,$

\vspace{4pt}
\begin{scriptsize}
\textsf{Q$_4$. \textbf{create table} Y\_s \textbf{as} \textbf{select} U.phi, U.upsilon, U.t, U.s \textbf{from}}\\
\indent\indent \textsf{(\textbf{select} phi, upsilon, t, s \textbf{from} Y1\_s \textbf{union all}}\\
\indent\indent \textsf{\textbf{select} phi, upsilon, t, s \textbf{from} Y2\_s \textbf{union all}}\\
\indent\indent \textsf{\textbf{select} phi, upsilon, t, s \textbf{from} Y3\_s) \textbf{as} U, Y\_Exp \textbf{as} E}\\
\indent\indent \textsf{\textbf{where} U.phi=E.phi \textbf{and} U.upsilon=E.upsilon;}
\end{scriptsize}

\vspace{4pt}
\begin{scriptsize}
\textsf{Q$_5$. \textbf{select} phi, upsilon, s, \textbf{conf()} \textbf{as} Prior \textbf{from} Y\_s \textbf{where} t=3}\\
\indent\indent \textsf{\textbf{group by} phi, upsilon, s \textbf{order by} Prior \textbf{desc};}
\end{scriptsize}

\begin{figure}[t]\scriptsize
\centering
\begingroup\setlength{\fboxsep}{3pt}
\colorbox{yellow!15}{%
   \begin{tabular}{c|c|c|c|>{\columncolor[gray]{0.97}}c||>{\columncolor[gray]{0.92}}c}
  \textsf{Y[$s$]} & $\phi$ & $\upsilon$ & $s$ & \textcolor{red}{\textsf{Prior}} & \textcolor{blue}{\textsf{Posterior}}\\
      \hline    
   & $1$ & $1$ & $2188.36$ & $.1$ & $.167$\\
   & $1$ & $1$ &  $2205.82$ & $.1$ & $.168$\\
   & $1$ & $1$ &  $2320.51$ & $.1$ & $.167$\\
   & $1$ & $1$ &  $2337.97$ & $.1$ & $.165$\\
   & $1$ & $1$ &  $2452.66$ & $.1$ & $.149$\\
   & $1$ & $1$ &  $2470.12$ & $.1$ & $.145$\\
\cline{2-6}
   & $1$ & $2$ & $2930.59$ & $.05$ & $.020$\\
   & $1$ & $2$ &  $2943.44$ & $.05$ & $.019$\\
   & $1$ & $2$ &  $4991.92$ & $.05$ & $.000$\\
   & $1$ & $2$ &  $4991.97$ & $.05$ & $.000$\\
\cline{2-6}
   & $1$ & $3$ &  $4778.87$ & $.05$ & $.000$\\
   & $1$ & $3$ &  $4779.56$ & $.05$ & $.000$\\
   & $1$ & $3$ &  $4944.72$ & $.05$ & $.000$\\
   & $1$ & $3$ &  $4944.89$ & $.05$ & $.000$ 
   \end{tabular}
}\endgroup\vspace{-7pt}  
\caption{$\Upsilon$-DB query for analytics on predicted position $s$.}\label{fig:analytics}
\vspace{-17pt}
\end{figure}

\noindent
pos\-sible $s$ tuples for \textsf{H}$_2$ and \textsf{H}$_3$ each. $\!$Considering all hypo\-theses $\upsilon$ for the same phenomenon $\phi$, the confidence values sum up to one in accordance with the laws of probability.

Users can make decisions in light of such confidence aggregates. $\!$\mbox{These are to be eventually conditioned in face of evi}\-dence (observed data). $\!$Example \ref{ex:bayes} features it for \emph{discrete} random variables mapped to the possible values of predictive attrib\-utes \mbox{(like position $s$) whose domain are \emph{continuous}}.
\vspace{-5pt}
\begin{myex}\label{ex:bayes}
Suppose position $s\!=\!2250$ feet is observed at \mbox{$t\!=\!3$} secs, with standard deviation $\sigma\!=\!20$. $\!\!$Then,$\!$ by applying Bayes' theorem for normal mean with a discrete prior \emph{\cite{bolstad2007}}, \textsf{Prior} is updated to \textsf{Posterior} (see Fig. \ref{fig:analytics}). $\Box$
\end{myex}
\vspace{-2pt}
\noindent
The procedure uses normal density function $\!$(\ref{eq:density}), $\!$with \mbox{$\sigma\!=\!20$}, to get the likelihood $f(y \,|\, \mu_i)$ of each alternative prediction of $s$ from \textsf{Y[$s$]} as mean $\mu_i$ given $y$ at observed $s\!=\!2250$. Then it applies Bayes' rule (\ref{eq:bayes}) to get the posterior $p(\mu_i \,|\, y)$.
\vspace{-2pt}
\begin{eqnarray}
f(y \,|\, \mu_i) \!\!&=&\!\! \frac{1}{\sqrt{2\pi \sigma^2}}\, e^{-\frac{1}{2\sigma^2}(y-\mu_i)^2}\label{eq:density}\\
p(\mu_i \,|\, y) \!\!&=&\!\! f(y \,|\,\mu_i)\;p(\mu_i) \;/\;\textstyle\sum_{i=1}^n f(y \,|\,\mu_i)\;p(\mu_i)\label{eq:bayes}
\end{eqnarray}

\noindent 

\vspace{-8pt}
\section{Research Challenges}\label{sec:challenges}

\vspace{-1pt}
\noindent
Big data in general, and hypotheses as data in particular, challenge traditional DB design methodologies 
\cite{badia2011}. $\!$Mean\-while, p-DB models like \textsf{MayBMS}' extend the relational mo\-del opening new opportunities, 
in particular for \emph{design by synthesis} \cite{bernstein1976}. New problems span from fast-varying schemas to uncertainty, probability and correlations in the raw data.

\textbf{Structural variety}. The user external view of the world is constantly changing. 
Our approach to this challenge consists in isolating or safeguarding alternative views under their own FD schemes and epistemological keys, allowing for their co-existance in the same p-DB in a controlled way. 

$\!$\textbf{Dependency extraction}. It has been considered a critical failure in traditional DB design the lack of techniques to obtain important information (e.g., FDs) in the real world \cite[$\!$p. $\!\!\!$62]{badia2011}. 
Synthesis methods for p-DB design shall provide novel abstractions and techniques to extract dependencies from other kinds of formal specification (e.g., equations).

\textbf{Schema synthesis}. $\!$Predictive data has correlations or, an \emph{uncertainty chaining}, we capture in FDs. Reasoning to synthesize relations has to account for that, viz., it has to go beyond 3NF and compute the pseudo-transitive closure (PTC) of each FD schema. For example, running the classical 3NF synthesis algorithm \cite{bernstein1976} on $\Sigma_1$ produces relation \mbox{\textsf{R$_i$($g$, $\!\operatorname{v_0}$, $\!s_0$, $\!\upsilon$, $\!t$, $\!s$)},} whereas we target at the also lossless, but less redundant \textsf{R$_i$($\!\phi$, $\!\upsilon$, $\!t$, $\!s$)}. That is, parameters are to be folded \emph{for certainty} (run PTC on $\Sigma_1$), and then unfolded \emph{for uncertainty} (re-run it on $\Sigma_1 \setminus \{ \phi \!\to\! g \operatorname{v_0} s_0 \}$).

$\!\!$\textbf{Uncertainty factors}. $\!$PTC for uncertainty (u-PTC) must synthesize prediction relations (e.g., $\!$\textsf{Y1[$a$]}) with the proper uncertainty factors in their condition columns. Besides $\upsilon$, a trivial factor, identifying each independent uncertainty unit from trials (cf. $\!$\textsf{H1\_INPUT}) with one, only one random variable is a combinatorial problem of \emph{uncertainty factor learning}. $\!$Thus, u-PTC must be sensitive not exactly to parameters $A$ but to the uncertainty factors $\vec{X}\!\subseteq\! \vec{A}$ they fall into.

$\!\!$\textbf{Cyclic FDs}. $\!$In the hypotheses of Ex. $\!$\ref{ex:fall}, no \mbox{prediction var}\-iable is dependent on each other. $\!$Complex mathematical models, however, have coupled variables leading to cyclic FDs like $\{a\,x\,\upsilon \!\to y,\, b\,y\,\upsilon \!\to x\}$. $\!$This is a specific issue of cycles in the uncertainty chaining for the (u-)PTC algorithm. 

$\!\!\!$\textbf{Conditioning}. The prior probability distribution assigned via \mbox{\textsf{repair $\!$key}} to$\!$ uncertainty factors (cf.$\!$ \textsf{Q}$_1$, $\!$\textsf{Q}$_2$) $\!$is to be eventually conditioned on observed data (Ex. $\!$\ref{ex:bayes}). \mbox{$\!$This is an ap}\-plied \emph{Bayesian inference} problem that translates into a \emph{p-$\!$DB update} one to induce effects of posteriors back to table $W\!$. $\!$It is achievable (yet unclean) in \textsf{MayBMS}' update language.

\vspace{-8pt}
\section{Conclusions}\label{sec:conclusions}
\vspace{-3.5pt}
\noindent
We have presented the vision of $\Upsilon$-DB, which is essentially an abstraction of hypotheses as uncertain data. It comprises a design methodology for the systematic construction and management of U-relational hypothesis DBs. To our knowledge this is the first design-by-synthesis method for constructing p-DBs from formal specifications.

$\!\!\!$We have introduced hypothesis management as a promising new class of applications for p-DBs, providing a principled approach to manage theoretical big data on top of \textsf{MayBMS}. 
The potential of $\,\Upsilon$-DB for deep predictive analytics has also been illustrated. $\!$First results are to be delivered from a large-scale use case in Computational Hemodynamics.

\nobalance
\vspace{-8.5pt}
\section{Acknowledgments}
\vspace{-3pt}
\noindent
This research has been supported by the Brazilian funding agencies CNPq, grants n$^o\!$ 141838/2011-6 and 309494/2012-5, and FAPERJ grant n$^o\!$ E-26/100.286/2013. $\!$We thank IBM for awarding this project a Ph.D. Fellowship 2013-2014. 

\vspace{-9pt}
\scriptsize
\bibliographystyle{abbrv}
\bibliography{vldb14}  

\begin{thebibliography}{10}

\bibitem{badia2011}
A.~Badia and D.~Lemire.
\newblock A call to arms: Revisiting database design.
\newblock {\em {SIGMOD} Record}, 40(3):61--9, 2011.

\bibitem{bernstein1976}
P.~Bernstein.
\newblock Synthesizing third normal form relations from functional
  dependencies.
\newblock {\em ACM TODS}, 1(4):277--98, 1976.

\bibitem{bolstad2007}
W.~M. Bolstad.
\newblock {\em Introduction to Bayesian Statistics}.
\newblock Wiley-Interscience, 2nd edition, 2007.

\bibitem{darwiche2010}
A.~$\!\!$Darwiche.
\newblock $\!${B}ayesian $\!$networks.
\newblock {\em $\!$Comm $\!$ACM}, $\!$53(12):80--90, $\!$2010.

\bibitem{sarma2007}
A.~{Das Sarma}, J.~Ullman, and J.~Widom.
\newblock Schema design for uncertain databases.
\newblock In {\em Proc. of AMW}, 2007.

\bibitem{haas2011}
P.~$\!$Haas, P.~$\!$Maglio, P.~$\!$Selinger$\!$, and W.~$\!$Tan.
\newblock $\!${Data} $\!$is $\!$dead... without what-if models.
\newblock {\em PVLDB}, 4(12):1486--9, 2011.

\bibitem{hey2009}
T.~Hey et~al.
\newblock {\em The Fourth Paradigm: Data-Intensive Scientific Discovery}.
\newblock Microsoft Research, 2009.

\bibitem{koch2009}
C.~Koch.
\newblock {\em May{BMS}: {A} system for managing large uncertain and
  probabilistic databases}.
\newblock In C. Aggarwal (ed.), Managing and Mining Uncertain Data, chapter 6.
  Springer-Verlag, 2009.

\bibitem{koch2008}
C.~Koch and D.~Olteanu.
\newblock Conditioning probabilistic databases.
\newblock {\em PVLDB}, 1(1):313--25, 2008.

\bibitem{losee2001}
J.~Losee.
\newblock {\em A historical introduction to the philosophy of science}.
\newblock Oxford University Press, 4th edition, 2001.

\bibitem{porto2011}
F.~Porto and S.~Spacappietra.
\newblock {\em Data model for scientific models and hypotheses}.
\newblock In R. Kaschek, L. Delcambre (ed.), The evolution of Conceptual
  Modeling, p. 285-305, Springer-Verlag LNCS vol. 6520, January 2011.

\bibitem{simon1966}
H.~Simon and N.~Rescher.
\newblock Cause and counterfactual.
\newblock {\em Philosophy of Science}, 33(4):323--40, 1966.

\bibitem{suciu2011}
D.~Suciu, D.~Olteanu, C.~R\'e, and C.~Koch.
\newblock {\em $\!$Probabilistic$\!$ Databases}.
\newblock Morgan \& Claypool Publishers, 2011.

\bibitem{ullman1988}
J.~Ullman.
\newblock {\em $\!$Principles $\!$of {D}atabases $\!$and {K}nowledge-{B}ase
  {S}ystems}.
\newblock Computer Science Press, 1988.

\end{thebibliography}


\end{document}